\documentclass[preprint,superscriptaddress,showpacs]{revtex4}
\usepackage{epsfig,amsmath,graphicx,amssymb}
\usepackage[colorlinks=fals,linkcolor=blue]{hyperref}
\usepackage{booktabs}
\usepackage{tabularx}
\usepackage{color}
\usepackage{dcolumn}
\usepackage{bm}
\usepackage{longtable}
\usepackage{CJK}
\def\be{\begin{equation}}
\def\ee{\end{equation}}
\def\bea{\begin{eqnarray}}
\def\eea{\end{eqnarray}}

\newcommand{\beq}{\begin{equation}}
\newcommand{\eeq}{\end{equation}}
\usepackage{graphicx}
\usepackage{dcolumn}
\usepackage{bm}
\usepackage{longtable}
\usepackage{color}
\usepackage{CJK}
\begin{document}
\begin{CJK*}{GBK}{song}
\title{Calculation of $\alpha$-decay energies of superheavy nuclei in a hybrid method}
\author{Z. Li}
\affiliation{School of Physics and Nuclear Energy Engineering, Beihang University, China, 100191.}
\affiliation{International Research Center for Nuclei and Particles in the Cosmos, Beijing 100191, China}
\author{B. Sun}
\email{bhsun@buaa.edu.cn}
\affiliation{School of Physics and Nuclear Energy Engineering, Beihang University, China, 100191. }
\affiliation{International Research Center for Nuclei and Particles in the Cosmos, Beijing 100191, China}
\author{C.H. Shen}
\affiliation{School of Physics and Nuclear Energy Engineering, Beihang University, China, 100191. }
\author{W. Zuo}
\email{zuowei@impcas.ac.cn}
\affiliation{Institute of Modern Physics, Chinese Academy of Sciences, China, 730000. }
\date{\today}

\begin{abstract}
A recent proposed method for $\alpha$-decay energies ($Q_\alpha$) [J.M. Dong, W. Zuo, and W. Scheid, Phys. Rev. Lett. \textbf{107}, 012501 (2011)] can reproduce experimental data of superheavy nuclei (SHN) with an $rms$-value of less than 100 keV. However, a sinusoid-like periodic deviation from experiments, which limits the accuracy in predictions, is observed when using different reference nuclei. In this paper, we have further extended this hybrid method, i.e., to predict $Q_\alpha$ of the as-yet-unobserved SHN with the help of known nuclei. It is found that the systematic deviation in previous study
is rooted in the nuclear mass model employed. By further analyzing the source of errors, different nuclear mass models are evaluated based on the same procedure.
\end{abstract}

\maketitle



In the exploration of the terra incognita of our nature, it has been an active and fascinating field for many years to search the mass and charge limits of realistic nuclei, i.e. the superheavy nuclei (SHN) \cite{Hofmann1998,Hofmann2000}.
For the SHN with atomic number beyond $110$ investigated so far, it has been revealed that the $\alpha$-decay is the dominant decay mode
and only 7 nuclei decay with spontaneous fission.
And for all the 41 SHN with $Z\geq107$ synthesized by Ca$^{48}$-induced reactions in Dubuna,
34 of them decay only by $\alpha$ emission while
only 7 of them undergo spontaneous fission \cite{Oganessian2013}.
Naturally, the $\alpha$-decay has long been a valuable tool for the identification of new elements via the observation of $\alpha$-decay chains from unknown nuclei to known nuclei when searching for SHN.
Precise measurements of $\alpha$-decay properties meanwhile provide rich information on the structure of SHN, ranging from ground-state energies, shell effects and stability, to nuclear spins and parities, nuclear deformation, and shape coexistence.
Accurate $\alpha$-decay $Q$ values ($Q_\alpha$)
are also crucial for predicting the half-lives of SHN.
In the nuclear astrophysical point of view, $\alpha$-decay of heavier elements and possibly SHN accounts for the interpretation of the third r-process abundance peak and therefore may affect the actinide chronometers~\cite{Cowan1999,Goriely01,Schatz02,Kratz07,Niu2009,Zhang2012a}.

In the last two decades, theoretical efforts in particular focusing on the predictions of $Q_\alpha$ (e.g. Ref.~\cite{buck,kumar,Zhang05,xu,zhang,pei,chowdhury,bhattacharya,ni,Zhao12})
for SHN had earned inspiring progresses.
Recently, a hybrid method is proposed according to the correlation between the
$\alpha$-decay energies of superheavy nuclei, aiming to predict the $\alpha$-decay energies of superheavy nuclei more accurately~\cite{Dong2011}. In this new approach, the $Q_{\alpha}$ value of any specific nucleus (target nucleus) can be derived from the nearby nuclei (reference nuclei) with experimentally determined $Q_\alpha$ values. It turns out that the predictive power can be significant enhanced. However, the calculated $Q_\alpha$ values suffer a global deviation from experimental ones by showing a sine-like oscillation (see Fig.1 in Ref.~\cite{Dong2011}) although small in amplitude, which was not explicated in the previous work. In the present work, the deviation is studied in detail and furthermore attributed to the uncertainties of nuclear mass model used. We extend this approach as well by using other widely used global mass models.

If shell effects {\it etc.} that can dramatically change the mass surface are not introduced, $Q_\alpha$ should change smoothly within the mass surface. On the plane of variables ($A$, $\beta$), where $A$ is the mass number and $\beta=(N-Z)/A$ the isospin asymmetry,
the $\alpha$-decay $Q$ value ($Q_2$) of the nucleus with ($A$, $\beta_2$) can be deduced in the first-order from
a known $\alpha$-decay $Q$ value ($Q_1^{\text{exp.}}$) of a nuclide with (A, $\beta_1$) by
\beq
\label{eq:eq1}
Q_2~\approx~ Q^{\text{exp.}}_1+(\beta_2-\beta_1)\bigg(\frac{\partial Q}{\partial \beta}\bigg).
\eeq
This formalism can be generalized further to the plane of ($\xi$, $\beta$) \cite{Dong2011},
where $\xi=xZ+yN$; $x$ and $y$ are both integers.

Eq. \eqref{eq:eq1} describes a short range correlation between $Q_{\alpha}$ values in the superheavy region,
which results from the local continuity of nuclear interactions. One can
reinterpret the second term in the right-hand-side of Eq. \eqref{eq:eq1} as
\beq
\begin{split}
\label{eq:eq2}
(\beta_2-\beta_1)\bigg(\frac{\partial Q}{\partial \beta}\bigg)&~\approx~(\beta_2-\beta_1)\bigg(\frac{Q_2^{\text{th.}}-Q_1^{\text{th.}}}{\beta_2-\beta_1}\bigg)\\
&~=~
Q_2^{\text{th.}}-Q_1^{\text{th.}} \;,
\end{split}
\eeq
in which $Q_1^{\text{th.}}$ and $Q_2^{\text{th.}}$ represent the
theoretically predicted $\alpha$-decay $Q$ values.
Instead of using the pure theoretic prediction, an unknown $Q$ value ($Q_2$) can be derived alternatively with
\beq
\label{eq:eq3}
Q_2~=~Q_1^{\text{exp.}}+(Q_2^{\text{th.}}-Q_1^{\text{th.}})
\eeq
The above prediction thus only relies on the finite differences in $Q$-values in theoretical models,
which generally can be predicted in a better accuracy owing to the cancelation of systematic errors.
Eq.~\eqref{eq:eq3} can be rewritten as
\beq
\label{eq:eq4}
Q_2~=~Q_2^{\text{th.}}+(Q_1^{\text{exp.}}-Q_1^{\text{th.}}) \;,
\eeq
revealing that once the target nucleus is chosen, the calculated $\alpha$-decay $Q_2$ value will
indeed share the same trend of the theoretical deviation from experimental data for the reference nucleus.

\begin{figure}
  \label{fig1}
  \includegraphics[width=0.8\textwidth]{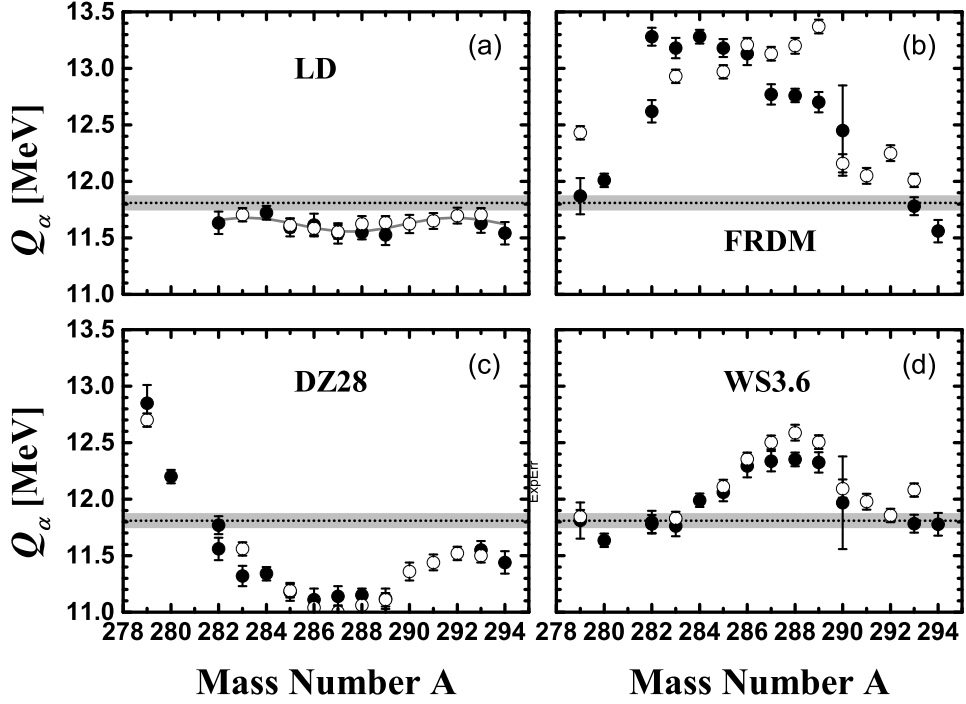}
  \caption{Deviations of $Q_\alpha$ calculated using Eq.~\eqref{eq:eq3} within various mass models with respect to the experimental $Q$ value (dot line with shaded area) for $^{294}$118. The open (filled) circles denote results calculated from the reference nuclei with even (odd) proton numbers, and the horizontal ordinate is the mass numbers of the reference nuclei. The uncertainties (error bars) of calculated values result from those $Q_\alpha$ of reference nuclei. In panel (a) the deviation from experimental $Q_\alpha$
  does behave like a sine-like pattern as shown by the grey solid line.}
\end{figure}

As an example, the SHN $^{294}$118 is taken to demonstrate the ability of the hybrid method, namely, Eq.~(\ref{eq:eq3}).
In Fig. 1, the predicted $Q_\alpha$ values by Eq.~(\ref{eq:eq3}) within various mass models are compared with the experimental data. The result based on the liquid-drop model (LD)~\cite{Dong2008} is shown in Fig.~1(a). However, being different from Ref.~\cite{Dong2011} where 380 reference nuclei were used to compute the $Q_\alpha$ values of target nuclides, here only 26 nuclides with proton number $Z\geqslant 110$ are used as reference nuclei.
The corresponding nuclide(s) for each mass number $A$ in the horizontal axes is(are) shown in Table I. The results indeed display a sine-like pattern with an amplitude of 0.0677 MeV and a wavelength of 8.809 mass units as shown by the grey solid line. Even more surprising is that the "same" periodic oscillation holds as well for all the calculated $Q_{\alpha}$ values of other SHN. This offers a way to get more accurate predictions towards unknown $Q_\alpha$ by examining the systematic deviation from experiments for well-known nuclei in the same mass region. As for $^{294}$118, the agreement between the experimental and theoretical values can be further improved from a $rms$ value of 0.196 MeV to 0.038 MeV after this periodic systematic error being subtracted.

\begin{table}
\caption{\label{tab1}%
List of reference nuclei used in this work. Shown in each line are the $Q_{\alpha}$ value and the literature for a certain isotope.
}
\begin{tabular}{l c c | l c c}
\hline\hline
 Isotope     & $Q_{\alpha}$ (MeV)& Reference                 & Isotope     & $Q_{\alpha}$ (MeV)& Reference                 \\
\hline
 $^{279}$110 & 9.84(6)           & Ref. \cite{Oganessian04a} & $^{288}$114 & 10.09(7)          & Ref. \cite{Oganessian04a} \\
 $^{279}$111 & 10.52(16)         & Ref. \cite{Oganessian04b} & $^{289}$114 & 9.96(6)           & Ref. \cite{Oganessian04a} \\
 $^{280}$111 & 9.87(6)           & Ref. \cite{Oganessian04b} & $^{287}$115 & 10.74(9)          & Ref. \cite{Oganessian04b} \\
 $^{282}$111 & 9.13(10)          & Ref. \cite{Oganessian10}  & $^{288}$115 & 10.61(6)          & Ref. \cite{Oganessian04b} \\
 $^{283}$112 & 9.67(6)           & Ref. \cite{Oganessian04a} & $^{289}$115 & 10.45(9)          & Ref. \cite{Oganessian10}  \\
 $^{285}$112 & 9.29(6)           & Ref. \cite{Oganessian04a} & $^{290}$115 & 10.09(40)         & Ref. \cite{Oganessian10}  \\
 $^{282}$113 & 10.78(8)          & Ref. \cite{Oganessian07}  & $^{290}$116 & 11.00(8)          & Ref. \cite{Oganessian04a} \\
 $^{283}$113 & 10.26(9)          & Ref. \cite{Oganessian04b} & $^{291}$116 & 10.89(7)          & Ref. \cite{Oganessian04a} \\
 $^{284}$113 & 10.15(6)          & Ref. \cite{Oganessian04b} & $^{292}$116 & 10.80(6)          & Ref. \cite{Oganessian04a} \\
 $^{285}$113 & 9.88(8)           & Ref. \cite{Oganessian10}  & $^{293}$116 & 10.67(6)          & Ref. \cite{Oganessian04a} \\
 $^{286}$113 & 9.76(10)          & Ref. \cite{Oganessian10}  & $^{293}$117 & 11.18(8)          & Ref. \cite{Oganessian10}  \\
 $^{286}$114 & 10.33(6)          & Ref. \cite{Oganessian04a} & $^{294}$117 & 10.96(10)         & Ref. \cite{Oganessian10}  \\
 $^{287}$114 & 10.16(6)          & Ref. \cite{Oganessian04a} & $^{294}$118 & 11.81(6)          & Ref. \cite{Oganessian04a} \\
\hline\hline
\end{tabular}
\end{table}

In Fig.1~(b-d) the deviations from experimental data are shown as well when employing the finite-range droplet model (FRDM)~\cite{Moller1995}, Weizs\"acker-Skyrme (WS3.6) mass model~\cite{Liu2011}, as well as the Duflo-Zuker (DZ28) mass model~\cite{Duflo1995}. The model-dependent deviations from experimental data are clearly seen, e.g. a parabola-like deviation for the cases with FRDM, WS3.6 and DZ28.
Meanwhile, the hybrid method can be repeated in each model to calculate other SHN. The $rms$ deviations in the new approach for FRDM, WS, DZ and LD with respect to experimentally determined values are 0.56 MeV, 0.27 MeV, 0.47 MeV and 0.31 MeV, respectively, while the $Q$ values predicted in the corresponding models are 0.65 MeV, 0.27 MeV, 0.77 MeV, 0.36 MeV, respectively.
The extended Thomas-Fermi plus Strutinsky integral with quenched shell (ETFSI-Q)~\cite{Pearson1996}, has also been examined, and the corresponding $rms$ value is reduced even by a factor of 2 from 0.53 MeV to 0.27 MeV by using the present method [Eq.(\ref{eq:eq3})].
Again here the comparisons include only nuclei heavier than darmstadium. However, the new approach does not improve the prediction of $Q_\alpha$ for WS3.6.

  \begin{figure}
      \includegraphics[width=0.8\textwidth]{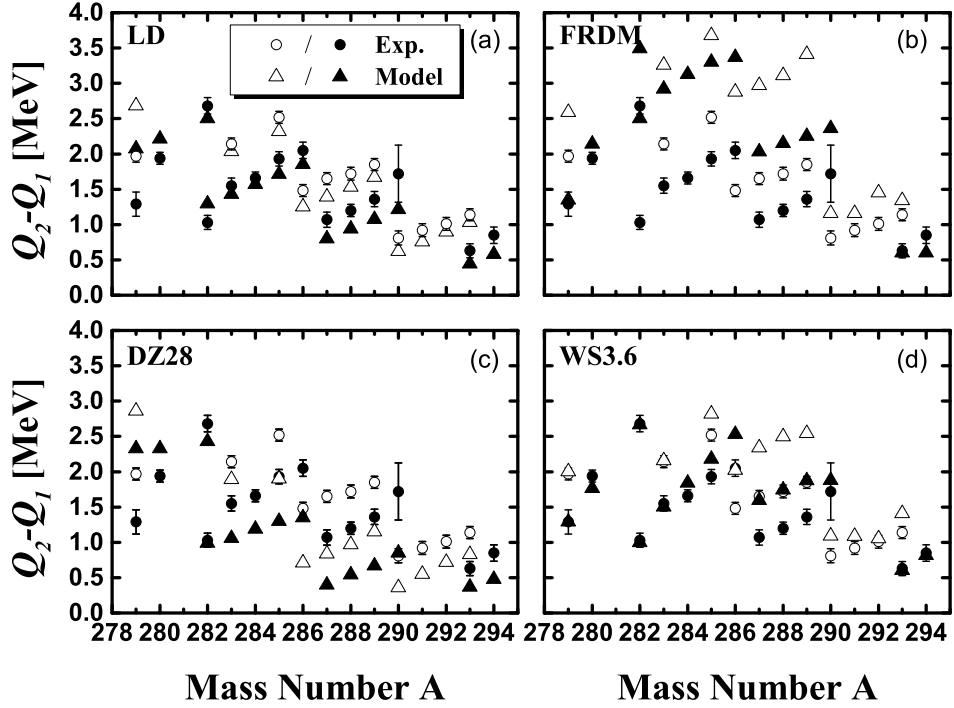}
      \caption{Experimental (circles) and theoretical (triangle) differences between the $Q_\alpha$ values of the target nucleus $^{294}118$ and
      reference nuclei. The open (filled) symbols indicate the values calculated using reference nuclei with even (odd) proton number.
      The corresponding nuclide(s) for each mass number $A$ in the horizontal axes is(are) shown in Table I.}
  \label{fig2}
 \end{figure}

To further illustrate the source of errors in Eq.~(\ref{eq:eq3}), the disparity $\delta Q$ between calculated and experimental $Q$ values can be given as
\beq
\begin{split}
\delta Q&~=~Q^{\text{exp.}}_2-Q_2\\
&~=~(Q^{\text{exp.}}_2-Q_1^{\text{exp.}})-(Q^{\text{th.}}_2-Q_1^{\text{th.}}),
\end{split}
\eeq
which attributes the disparity to the difference between the experimental and theoretical parts. Again
let us take the SHN $^{294}118$ as an example. Given the same set of reference nuclei, the experimental and the theoretical differences between the $Q_{\alpha}$ values of the target nucleus and reference nuclei, i.e. $Q_2-Q_1$, are shown in Fig.~\ref{fig2}, respectively.
It is seen that the FRDM and WS3.6 models overestimate the $Q_\alpha$ differences, whereas the DZ28 case favor a weak evolution in the $Q_\alpha$ surface. The LD case fits the experimental data best, therefore leading a more faithful description of the $Q_{\alpha}$ values of $^{294}118$.
However, it is worth noticing that the LD model used here has its parameters optimized locally for the superheavy region \cite{Dong2010}. Naturally, the dedicated optimization tends to provide a better description than those global models. It is extremely interesting to examine the ability of the hybrid method with microscopic-rooted models like the covariant density functional approaches. Unfortunately, up to now the mass values from these models are rather rare in literatures.

In Fig.3, we compare the theoretical $Q$ values obtained from Eq.~\eqref{eq:eq3}
and the corresponding experimental ones. The value for each single nucleus is calculated by averaging all the results
computed from reference nuclei with $Z\geqslant 110$. In comparison with the $Q_\alpha$ values directly calculated from the DZ28 model, better agreements are clearly obtained by the hybrid method, especially for the isotopic chains of 116 and 117 where more reference nuclei can be used.
The predictive power of $Q_\alpha$ in the hybrid method, in principle, could be further improved by adding more reference nuclei. Reliable Q-values predicted in this work combined with directly measured masses allows one to determine the masses of these nuclides linked with $\alpha$-decays and, hence, to extend the measured mass-surface considerably to the exotic region of nuclides.

  \begin{figure}
  \label{fig3}
      \includegraphics[width=0.8\textwidth]{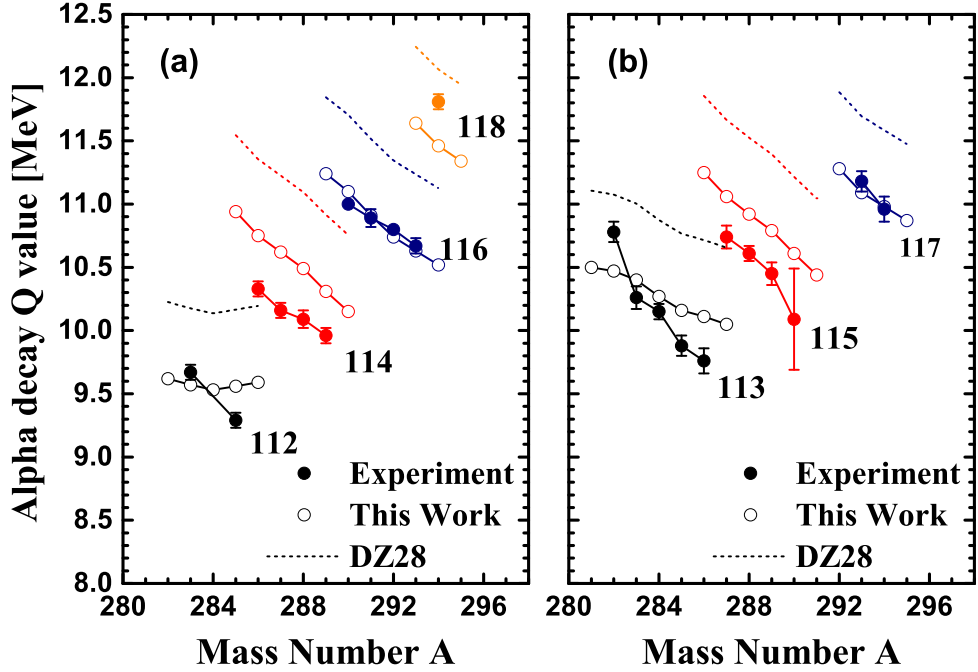}
      \caption{(Color online) The comparison of experimental $\alpha$-decay $Q$ values (filled circle) and theoretical values using the present method (open circle). The $Q_\alpha$ values in the DZ28 model are shown by the dash lines. The even-Z and odd-Z isotopic chains are plotted in the panel (a) and (b), respectively. The various colors for isotopic chains are used only to guide the eye.}
 \end{figure}

In summary, the sinusoid-like periodic deviation from experiments in Ref.~\cite{Dong2011}, in which a new hybrid method is proposed to predict $\alpha$-decay energies, is found to be rooted in the nuclear mass model employed. By removing such periodic patterns, one can achieve even
better predictive accuracy down to about 40 keV as demonstrated by the case of SHN $^{294}118$. The accuracy of the new approach
has been then traced back to the evolution of $Q_\alpha$, which in turn limits the accuracy in predictions. More reference nuclei employed from experimental result could improve further the reliability in prediction.
Practically, this offers a way to reliable prediction of the $Q_\alpha$ values of the as-yet-unobserved SHN with the help of known nuclei.
It is also very interesting to extend the new approach to microscopic-rooted models like the covariant density functional theories,
since these approaches have shown better reproductive power in derivatives of absolute masses such as one-neutron separation energies $S_n$~\cite{Sun2008} and Coulomb displacement energies~\cite{Brown2002,Sun2011}. Those effects such as pairing correlations that are hard to be treated can be removed to a large extent in the hybrid approach.

The work is supported by the National Natural Science
Foundation of China (11035007, 11105010, 11128510, 11235002 and 11175219), the Major State Basic
Research Developing Program of China under No. 2013CB834405, the NECT under Grant No. NCET-09-0031 and
the Knowledge Innovation Project (KJCX2-EW-N01) of Chinese Academy of Sciences.

\end{CJK*}

\begin{thebibliography}{99}


\bibitem{Hofmann1998}
S. Hofmann, Rep. Prog. Phys. \textbf{61}, 639 (1998).
\bibitem{Hofmann2000}
S. Hofmann and G. M\"unzenberg, Rev. Mod. Phys. \textbf{72}, 733 (2000).
\bibitem{Oganessian2013}
Yu. Ts. Oganessian, Nuclear Physics News \textbf{23}:1, 15 (2013).
\bibitem{Cowan1999}
J. J. Cowan, B. Pfeiffer, K.-L. Kratz, F.-K. Thielemann, C. Sneden, S. Burles, D. Tytler, and T. C. Beers, Astrophys. J. \textbf{521}, 194 (1999).
\bibitem{Goriely01}
S. Goriely and M. Arnould, Astron. Astrophys. \textbf{379}, 1113 (2001).
\bibitem{Schatz02} H. Schatz, R. Toenjes, B. Pfeiffer, T. C. Beers, J. J. Cowan,
V. Hill, and K.-L. Kratz, Astrophys. J. \textbf{579}, 626 (2002).
\bibitem{Kratz07} K.-L. Kratz, K. Farouqi, B. Pfeiffer, J. W. Truran, C. Sneden,
and J. J. Cowan, Astrophys. J. \textbf{662}, 39 (2007).
\bibitem{Niu2009}
Z. Niu, B. Sun, and J. Meng, Phys. Rev. C \textbf{80}, 065806(2009).
\bibitem{Zhang2012a}
W. Zhang, Z. Niu, F. Wang, X. Gong, and B. Sun, Acta Phys. Sin. \textbf{61}, 112601(2012).
\bibitem{buck} B. Buck, A. C. Merchant, and S. M. Perez, Phys. Rev. Lett. \textbf{72}, 1326 (1994).
\bibitem{kumar} S. Kumar, M. Balasubramaniam, R. K. Gupta, G. Munzenberg, and W. Scheid, J. Phys. G: Nucl. Part. Phys., \textbf{29}, 625 (2003).
\bibitem{Zhang05} W. Zhang, J. Meng, S.Q. Zhang, L.S. Geng, H. Toki, Nucl. Phys. A \textbf{753}, 106 (2005).
\bibitem{xu} C. Xu, Z. Ren, Phys. Rev. C \textbf{73}, 041301(R) (2006).
\bibitem{zhang} H. F. Zhang, W. Zuo, J. Q. Li, and G. Royer, Phys. Rev. C \textbf{74}, 017304 (2006); G. Royer and H. F. Zhang, Phys. Rev. C \textbf{77}, 037602 (2008).
\bibitem{pei} J. C. Pei, F. R. Xu, Z. J. Lin, and E. G. Zhao, Phys. Rev. C \textbf{76}, 044326 (2007).
\bibitem{chowdhury} C. Samanta, P. Roy Chowdhury, and D. N. Basu,  Nucl. Phys. A \textbf{789}, 142 (2007); P. Roy Chowdhury, C. Samanta, and D. N. Basu, Phys. Rev. C \textbf{77}, 044603 (2008).
\bibitem{bhattacharya} M. Bhattacharya and G. Gangopadhyay, Phys. Rev. C \textbf{77}, 047302 (2008).
\bibitem{ni} D. D. Ni and Z. Z. Ren, Phys. Rev. C \textbf{81}, 064318 (2010).

\bibitem{Zhao12}
H. Jiang, G.J. Fu, B. Sun et al., Phys. Rev. C \textbf{85}, 054303 (2012)
\bibitem{Dong2011}
J. M. Dong, W. Zuo, and W. Scheid, Phys. Rev. Lett. \textbf{107}, 012501 (2011).
\bibitem{Dong2008}
T. Dong and Z. Ren, Phys. Rev. C \textbf{77}, 64310 (2008).
\bibitem{Oganessian04a}
Yu. Ts. Oganessian et al., Phys. Rev. C \textbf{70}, 064609 (2004).
\bibitem{Oganessian04b}
Yu. Ts. Oganessian et al., Phys. Rev. C \textbf{69}, 021601(R) (2004).
\bibitem{Oganessian10}
Yu. Ts. Oganessian et al.,Phys. Rev. Lett. \textbf{104}, 142502 (2010).
\bibitem{Oganessian07}
Yu. Ts. Oganessian et al., Phys. Rev. C \textbf{76}, 011601(R) (2007).
\bibitem{Moller1995}
P. M\"oller, J. Nix, W. Myers, and W. Swiatecki, At. Data Nucl. Data Tables \textbf{59}, 185 (1995).
\bibitem{Liu2011}
M. Liu, N. Wang, Y. Deng and X. Wu, Phys. Rev. C \textbf{84}, 014333 (2011).
\bibitem{Duflo1995}
J. Duflo, A. P. Zuker, Phys. Rev. C \textbf{52}, 23 (1995).
\bibitem{Pearson1996}
J. M. Pearson, R. C. Nayak, and S. Goriely, Phys. Lett. B \textbf{387}, 455 (1996).
\bibitem{Dong2010}
J. M. Dong, W. Zuo, J. Z. Gu, Y. Z. Wang, and B. B. Peng, Phys. Rev. C \textbf{81}, 064309 (2010).
\bibitem{Sun2008}
B. Sun, F. Montes, L. S. Geng, H. Geissel, Y. A. Litvinov, and J. Meng, Phys. Rev. C \textbf{78}, 025806 (2008).
\bibitem{Brown2002}
 B.A. Brown, R.R.C. Clement, H. Schatz, A. Volya and W.A. Richter, Phys. Rev. C \textbf{65}, 045802 (2002).
\bibitem{Sun2011}
B. Sun, P. Zhao, and J. Meng, Sci. China: Phys. Mech. Astron. \textbf{54}, 210 (2011).

\end{thebibliography}
\end{document}